# Design and Evaluation of Secure Distributed Chaotic Image Encryption on ESP32-S3

Mohammed Almozaini[1] ORCID
471009866@stu.iu.edu.sa
Fazal Noor[1] ORCID
Sohail Alhazmi[1] ORCID

[1]Faculty of Computer and Information Systems (FCIS), Islamic University of Madinah, Madinah 42351, Saudi Arabia

**Abstract:** The limited computational resources of Internet of Things (IoT) devices present challenges for secure image processing. This paper presents a distributed chaotic image encryption system using independent Wi-Fi-connected ESP32-S3 microcontrollers. A master-worker architecture distributes image blocks through master-coordinated chaotic sequence management, maintaining one logically continuous chaotic sequence by assigning block-specific starting states independently of worker assignments. The implemented system uses reversible chaotic substitution (SUBSTITUTE), while Elliptic Curve Digital Signature Algorithm (ECDSA)-signed discovery messages and Transport Layer Security (TLS) 1.3 with mutual authentication secure device communication. Experimental evaluation using one, two, and four workers demonstrates a reduction in execution time from 43.215 to 15.878 seconds, achieving a speed-up of 2.72×. The encrypted image achieves a Shannon entropy of 7.999940 bits/byte and near-zero adjacent-pixel correlation. Exact image recovery is verified through matching Secure Hash Algorithm 256-bit (SHA-256) hashes and a Structural Similarity Index (SSIM) value of 1.0. Additional experiments demonstrate successful encryption and decryption using different worker counts and block sizes, automatic task reassignment following worker failure, and rejection of a mismatched private key during TLS authentication. These findings demonstrate the feasibility of secure, flexible, and distributed chaotic image processing across independent resource-constrained IoT devices.

**Keywords**: Chaotic Encryption, Distributed Computing, Distributed Image Encryption, ECDSA, ESP32-S3, Internet of Things, TLS 1.3.

## 1. Introduction

The increasing use of Internet of Things (IoT) devices has created a growing demand for secure and efficient image processing at the network edge. Images generated by embedded devices may contain sensitive information requiring protection during storage and transmission. However, resource-constrained microcontrollers have limited processing power and memory, making computationally intensive image encryption challenging, particularly for larger images.

The ESP32-S3 was selected as a low-cost microcontroller with integrated Wi-Fi and limited computational resources. Unlike more powerful Raspberry Pi platforms, it represents a resource-constrained IoT environment. Potential applications include wireless security cameras, distributed surveillance systems, and edge-based image protection, where multiple devices cooperate to process image data.

Chaotic image encryption employs complex sequences for image permutation and substitution. Parallel and distributed processing techniques can address computational limitations by dividing workloads into smaller tasks [1], [2]. Previous studies have explored parallel processing environments, heterogeneous networks, processing architectures, and distributed execution techniques [3]–[6], providing a foundation for distributing chaotic image-processing workloads across independent devices.

Despite these developments, distributing the chaotic encryption workload of a single image across independent Wi-Fi-connected microcontrollers presents challenges in task coordination, chaotic sequence consistency, communication overhead, and device authentication. In particular, different worker assignments and image partitions must preserve the chaotic sequence required for correct decryption.

This paper presents an implemented master-coordinated distributed chaotic image encryption system using independent Wi-Fi-connected ESP32-S3 microcontrollers. A Python-based master partitions a single image into blocks, assigns block-specific chaotic starting states, distributes tasks among available workers, and coordinates result reconstruction. The implementation uses reversible chaotic substitution (SUBSTITUTE), ECDSA-signed discovery messages, and TLS 1.3 with mutual authentication for secure communication.

The main contributions of this work are:

1. A distributed architecture that processes a single image across independent ESP32-S3 workers using master-coordinated chaotic sequence management.
2. Reversible chaotic substitution supporting different worker counts and block sizes during encryption and decryption.
3. Integration of ECDSA authentication, TLS 1.3 communication, and automatic task reassignment following worker failure.
4. Experimental evaluation of scalability, image correctness, statistical characteristics, and communication security.

The experimental results demonstrate reduced execution time with additional workers, successful task recovery following worker interruption, and exact image reconstruction despite different encryption and decryption configurations.

## 2. Related Work

Parallel and distributed image processing has been widely investigated to overcome the computational limitations of processing large images on a single device. Prajapati and Vij [1] examined parallel and distributed image-processing approaches, while Kumaran and Quinn [2] explored divide-and-conquer programming for parallel computers. Subsequent studies introduced algorithmic frameworks, heterogeneous computing networks, and parallel processing architectures to improve workload distribution and execution efficiency [3]–[5]. Other researchers developed distributed processing backends, visual retrieval systems, medical imaging support systems, and embedded smart-camera middleware, demonstrating the feasibility of coordinating image-processing tasks across multiple computing resources [6]–[9].

Researchers also investigated hierarchical distributed frameworks and high-performance parallel processing techniques [10], [11]. These approaches were applied to ultra-high-definition imaging, edge detection, medical image processing, JPEG decoding, and hardware-accelerated convolution [12]–[16]. Additional distributed frameworks addressed satellite imagery, large medical datasets, remote sensing, and big-data image analysis [17]–[20]. Hristov et al. [21] demonstrated image-processing functionality through a web-based segmentation system. Although these studies establish the broader applicability of distributed image processing, their hardware platforms and computational objectives differ from distributed chaotic encryption on resource-constrained microcontrollers.

More directly related to this work, Raju et al. [22] distributed blocks of large images across a physical Hadoop cluster, while Che et al. [23] employed MPI-based parallel processing for image enhancement. These studies demonstrate image partitioning and distributed execution but do not establish coordinated distribution of chaotic-sequence segments alongside image blocks.

In image security, Ge et al. [24] proposed a chaotic encryption algorithm using divide-and-conquer diffusion. Other studies investigated hyperchaotic cryptosystems, parallel chaotic encryption, chaos-based steganography, and parallel computing for information protection [25]–[28]. Ping et al. [26] implemented parallel chaotic encryption within a single computer. Khan et al. [29] developed a feature-aware chaotic encryption algorithm incorporating pixel segmentation, permutation, and confusion. Additional research addressed image security in IoT environments, medical applications, and secure image transmission [30]–[32].

Abutaha et al. [33] developed an MPI-based parallel chaotic image encryption approach on multicore computers. Flores-Vergara et al. [34] implemented chaotic image-block encryption across six Raspberry Pi 3 systems connected through Ethernet, using MPI and dynamically generated keys for parallel encryption processes. Their work demonstrates chaotic encryption across physical embedded systems but does not establish master-coordinated partitioning of one logically continuous chaotic sequence into segments assigned independently of worker identity. Authenticated task reassignment and encryption consistency following worker failures were also not investigated.

Distributed image security has additionally been examined through cloud-based medical image encryption, homomorphic cryptography, and secure medical record storage [35]–[37]. Dinesh Kumar et al. [37] proposed a cloud-oriented DCUNA/CUNA framework that distributes medical image segments across simulated nodes using node-specific encryption keys and access-control policies. Although the framework incorporates authentication and fault tolerance, its implementation does not demonstrate task recovery across physical Wi-Fi-connected microcontrollers.

Bencherqui et al. [38] implemented a parallel medical image compression-encryption scheme on a four-node Raspberry Pi cluster, distributing compression, encryption, and optimization operations. Unlike their approach, the present work distributes the same chaotic substitution operation across independent ESP32-S3 workers, with each image block associated with its corresponding segment of a master-coordinated chaotic sequence.

TABLE 1
COMPARISON WITH RELATED STUDIES

| Study | Difference from proposed research |
|---|---|
| [22] Raju, 2017 | Hadoop image blocks; no chaotic encryption or wireless MCUs. |
| [23] Che, 2012 | MPI image enhancement; no chaotic encryption or Wi-Fi MCUs. |
| [24] Ge, 2022 | Algorithm-level diffusion; no physical distribution. |
| [26] Ping, 2020 | Parallel image encryption within one computer; no distributed physical workers. |
| [28] Zhang, 2023 | Parallel information protection; no demonstrated ESP32-S3 distributed chaotic processing. |
| [29] Khan, 2025 | Feature-aware chaotic encryption; no distributed computation across independent devices. |
| [33] Abutaha, 2022 | MPI-based chaotic encryption on multicore computers; no independent wireless workers. |
| [34] Flores-Vergara, 2019 | Six Raspberry Pi 3 systems using MPI over Ethernet for chaotic image-block encryption with dynamically generated per-process keys; master-coordinated partitioning of one continuous chaotic sequence and authenticated task reassignment are not established. |
| [37] Kumar, 2026 | Cloud-oriented distributed image encryption with unique keys per simulated node; no demonstrated master-coordinated chaotic processing on physical MCUs. |
| [38] Bencherqui, 2026 | Four-node Raspberry Pi cluster distributing compression, chaotic encryption, and optimization tasks; no demonstrated identical chaotic operation across independently assigned image blocks. |
| **This work** | **Master-coordinated image and chaotic-sequence partitioning; independent Wi-Fi ESP32-S3 workers; ECDSA authentication; TLS 1.3; automatic task reassignment.** |

The reviewed literature demonstrates substantial progress in parallel image processing, distributed image encryption, and embedded computing. However, previous studies have not demonstrated the distribution of a single image's chaotic encryption workload across independent Wi-Fi-connected microcontrollers while preserving one logically continuous chaotic sequence, regardless of worker assignments and image block sizes.

The implemented ESP32-S3 system addresses this gap through master-coordinated chaotic sequence management, authenticated communication, and automatic task reassignment following worker failure.

Direct performance comparisons are not made because the reviewed studies use different hardware platforms, algorithms, and communication architectures. A detailed feature-based comparison is provided in Appendix D, Table D1.

## 3. System Architecture and Implementation

### 3.1 System Architecture

The implemented system follows a master-worker architecture consisting of a Mac computer as the master, a PC server for image storage and transfer, and multiple Wi-Fi-connected ESP32-S3 microcontrollers as processing workers, as illustrated in Figure 1. ESP32-S3 devices were selected to investigate distributed image encryption using resource-constrained IoT hardware rather than more powerful platforms such as Raspberry Pi computers.

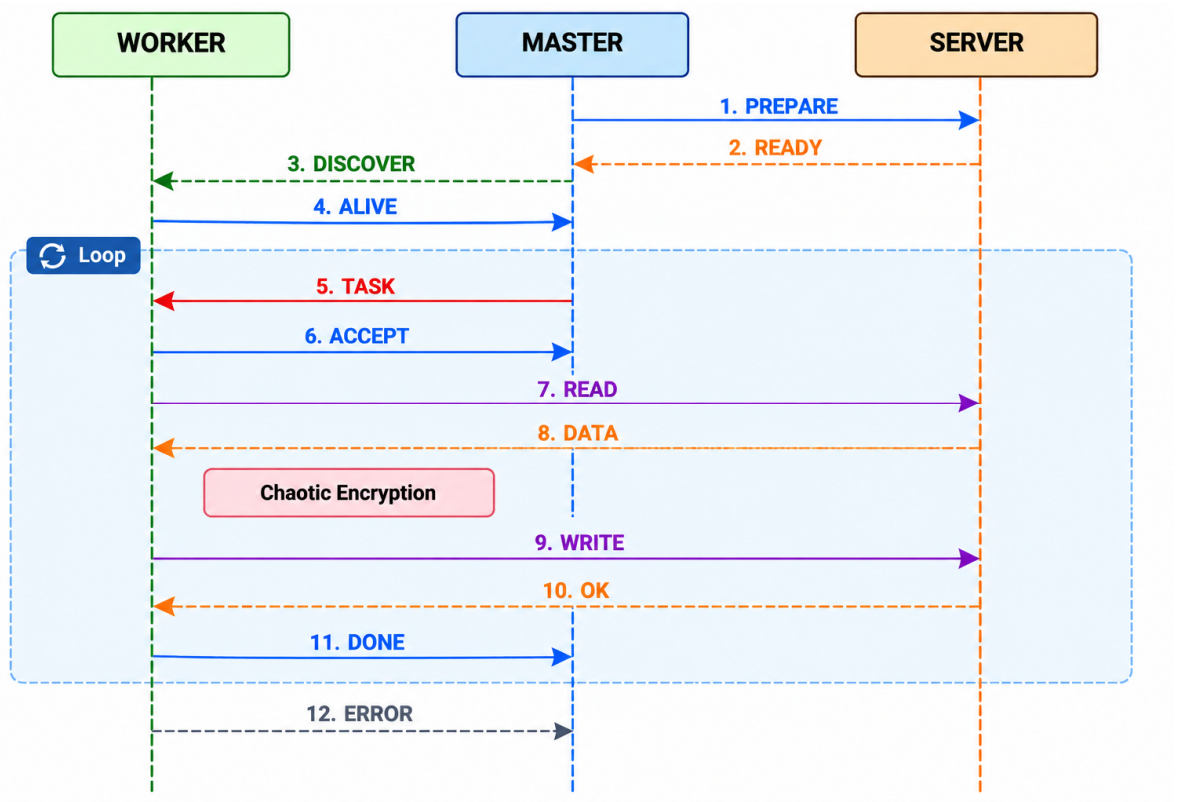

*Figure 1. Authenticated Distributed Chaotic Image Processing Architecture.*

The master initiates an image-processing session through the PC server, which prepares the image and returns its metadata. The master then discovers and authenticates available ESP32-S3 workers using ECDSA signatures based on the NIST P-256 (secp256r1) curve.

The master divides the image into blocks and assigns each block a starting state derived from one logically continuous chaotic sequence, independently of worker identity. This enables distributed processing while maintaining sequence continuity across different worker assignments and block sizes. The PC server transfers image blocks to the assigned workers and receives their processed results. Each worker performs the implemented chaotic substitution operation and notifies the master upon completion. Finally, the server reconstructs the processed blocks into the output image.

### 3.2 Chaotic Image Processing

The implemented system uses the SUBSTITUTE algorithm to modify pixel values through chaotic sequences. Workers independently generate their assigned chaotic sequence segments to process image blocks. The algorithm supports reversible encryption and decryption, as verified experimentally through exact image recovery. Future work may explore additional chaotic algorithms, including permutation, and their integration into combined encryption schemes.

### 3.3 Distributed Task Execution

The master divides the input image into configurable row blocks and distributes tasks among available ESP32-S3 workers. Each task follows five steps:

1. **Assignment:** The master assigns an image block and its chaotic starting state.
2. **Retrieval:** The worker retrieves the block from the BMP server.
3. **Processing:** The worker applies the SUBSTITUTE algorithm.
4. **Transfer:** The worker returns the processed block.
5. **Notification:** The worker informs the master of completion.

Task-specific chaotic starting states are independent of worker identity, supporting different worker counts, block sizes, and automatic task reassignment without affecting correctness. When a worker fails to complete an assigned task, the master can reassign it to another available worker using the same block-specific chaotic starting state.

### 3.4 Authentication and Communication Security

The system uses ECDSA digital signatures to authenticate UDP discovery messages, with sequence numbers (SEQ) providing replay protection. TCP communication is secured using TLS 1.3 with mutual authentication (mTLS), employing X.509 v3 certificates, P-256 public keys, and ECDSA-SHA256 signatures. These mechanisms protect device authentication and communication independently of the chaotic image-encryption algorithm. Authentication and secure communication were verified experimentally, including rejection of an invalid TLS certificate configuration.

## 4. Experimental Setup and Evaluation Metrics

### 4.1 Experimental Setup

The experiments were conducted using a Python-based master, a PC image server, and independent Wi-Fi-connected ESP32-S3 workers. A single RGB image was processed using the SUBSTITUTE algorithm. The experimental configurations and communication mechanisms are summarized in Table 2.

TABLE 2
EXPERIMENTAL CONFIGURATION

| Parameter | Configuration |
|---|---|
| Image | test.bmp |
| Resolution | 1280 × 855 pixels |
| Color format | 24-bit RGB |
| Processing algorithm | SUBSTITUTE |
| Task size | 128 and 256 image rows |
| Total tasks | 7 tasks (128 rows); 4 tasks (256 rows) |
| Worker configurations | 1, 2, and 4 ESP32-S3 devices |
| Scalability evaluation | 1, 2, and 4 workers using 128-row blocks |
| Worker flexibility | Encryption: 4 workers, 128 rows; decryption: 2 workers, 256 rows |
| Communication | Wi-Fi, UDP discovery, TLS 1.3 over TCP |
| Device authentication | ECDSA-signed UDP discovery; TLS 1.3 mTLS with X.509 v3 certificates, MAC-based CNs, P-256 keys, and ECDSA-SHA256 |

### 4.2 Evaluation Metrics

The system was evaluated using execution time, speed-up, and efficiency to assess processing performance and scalability. Execution time was measured from the first task assignment to final completion, including communication and authentication overhead but excluding initial discovery. Communication overhead represents data and message transfer costs, while authentication overhead refers to ECDSA and TLS 1.3 operations.

Image correctness was assessed using SHA-256 to verify exact recovery of the original image. Shannon entropy measured pixel-value randomness, adjacent-pixel correlation assessed statistical relationships, and the Structural Similarity Index (SSIM) evaluated image similarity.

### 4.3 Evaluation Equations

The number of image-processing tasks is calculated as:

$$N = \left\lceil \frac{H}{R} \right\rceil \quad (1)$$

where $H$ is the image height and $R$ is the number of rows per task.

Speed-up is calculated as:

$$S(n) = \frac{T_1}{T_n} \quad (2)$$

where $T_1$ and $T_n$ represent execution times using one and $n$ workers, respectively.

Efficiency is calculated as:

$$E(n) = \frac{S(n)}{n} \times 100\% \quad (3)$$

where $E(n)$ is parallel efficiency, $S(n)$ is speed-up, and $n$ is the number of ESP32-S3 workers.

Shannon entropy is calculated as:

$$H(X) = -\sum_{i=0}^{255} P(i) \log_2 P(i) \quad (4)$$

where $P(i)$ represents the probability of intensity value $i$. The theoretical maximum for 8-bit data is 8 bits/byte.

Pixel correlation is calculated as:

$$r_{xy} = \frac{Cov(x,y)}{\sqrt{D(x)D(y)}} \quad (5)$$

where *x* and *y* represent adjacent pixel values, *Cov(x,y)* is their covariance, and *D(x)* and *D(y)* are their variances. Correlation is evaluated horizontally, vertically, and diagonally.

Structural Similarity Index (SSIM) is calculated as:

$$SSIM(x,y) = \frac{(2\mu_x\mu_y + C_1)(2\sigma_{xy} + C_2)}{(\mu_x^2 + \mu_y^2 + C_1)(\sigma_x^2 + \sigma_y^2 + C_2)} \quad (6)$$

where $\mu_x$ *and* $\mu_y$ are the mean pixel values, $\sigma_x^2$ *and* $\sigma_y^2$ are the variances, $\sigma_{xy}$ is the covariance, and $C_1$ and $C_2$ are stability constants.

### 4.4 Evaluation Scope

The completed experiments assess execution time, speed-up, efficiency, image correctness, statistical characteristics, worker flexibility, fault tolerance, and communication security.

Communication and authentication overhead are included in the measured execution time but have not yet been separately quantified. The current evaluation is limited to one test image and single-run measurements for each configuration. Repeated trials, additional image resolutions, and detailed overhead measurements remain future work.

## 5. Results and Discussion

### 5.1 Execution Time and Scalability

The experiments evaluated distributed chaotic image substitution across different ESP32-S3 worker configurations. All scalability configurations used the same task size. Table 3 summarizes execution time, speed-up, and efficiency.

TABLE 3
DISTRIBUTED PROCESSING PERFORMANCE

| workers | Time (s) | Speed-up | Efficiency |
|---|---|---|---|
| 1 | 43.215 | 1.00× | 100.0% |
| 2 | 28.99 | 1.49× | 74.5% |
| 4 | 15.88 | 2.72× | 68.04% |

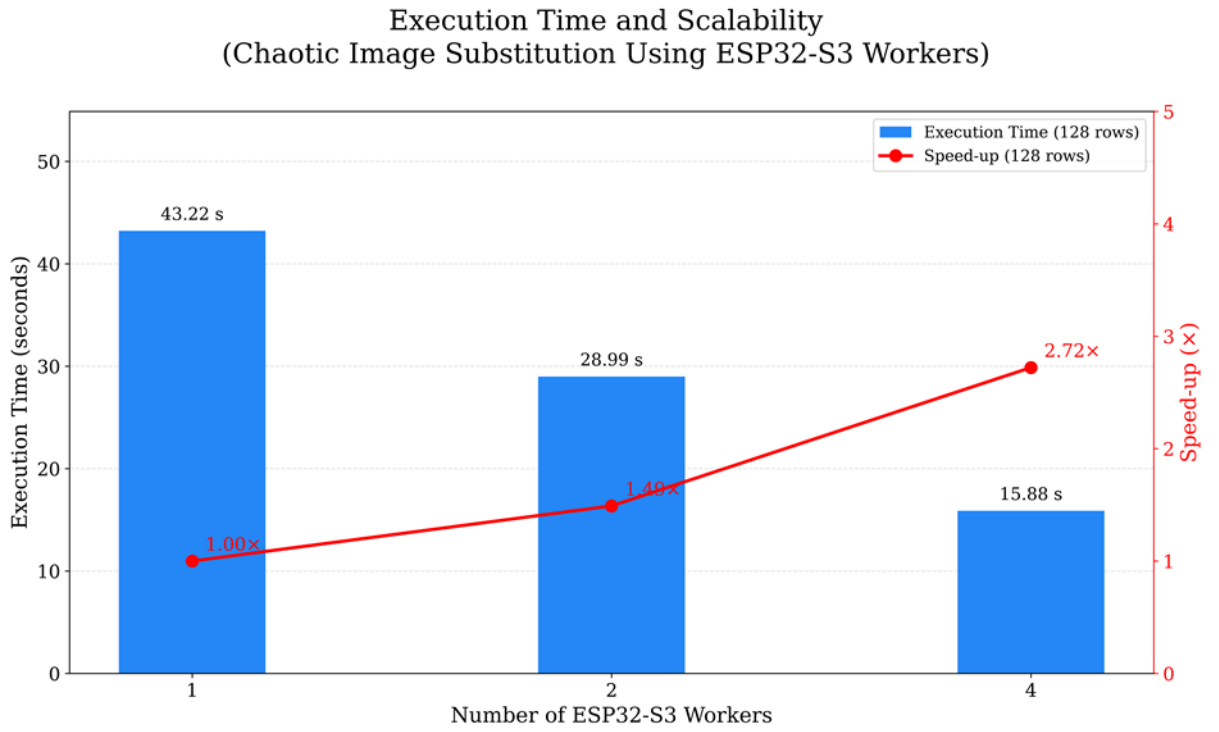

*Figure 2. Execution Time and Speed-up*

The results demonstrate that increasing the number of workers reduces execution time, as illustrated in Figure 2. However, efficiency decreases as the worker count increases, reflecting the effects of communication, coordination, and workload distribution overhead.

### 5.2 Fault-Tolerance Evaluation

A controlled fault-injection experiment was conducted to evaluate the system's ability to recover from worker failure during distributed image processing. An ESP32-S3 worker was deliberately interrupted after accepting an assigned image block, simulating a failure during task execution.

The master detected the incomplete task through its timeout mechanism, returned it to the pending queue, and automatically reassigned it to another available worker. Crucially, the master preserved the task's original chaotic starting state, allowing the replacement worker to reproduce the same chaotic-sequence segment independently of the failed worker.

This maintained chaotic-sequence continuity and ensured that worker failure did not affect the correctness of the reassigned block. The replacement worker successfully processed the block and returned the result to the BMP server, allowing the entire image to be completed without manual intervention.

As illustrated in Figure 3, the experiment demonstrates that automatic task reassignment can preserve both processing continuity and chaotic-sequence correctness despite an individual worker failure. The complete execution log is provided in Appendix A.

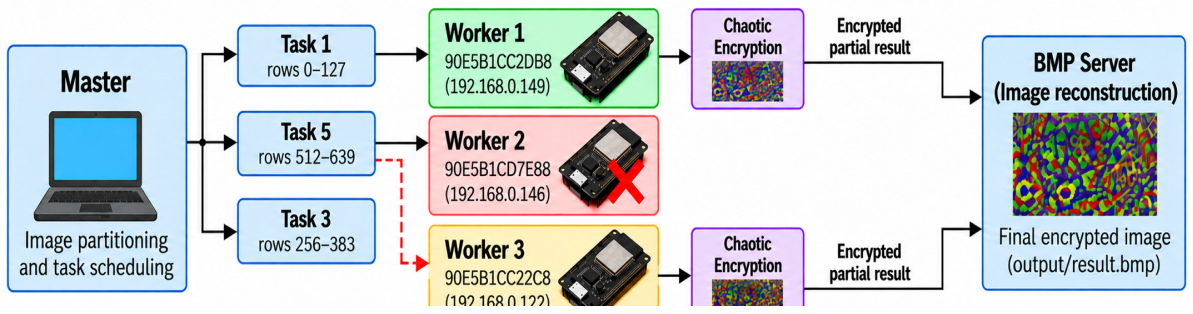

*Figure 3. Illustration of the fault-tolerance mechanism.*

### 5.3 Authentication and Communication Security

The experiments confirmed successful verification of ECDSA-signed UDP discovery messages and secure communication using TLS 1.3.

All assigned tasks were completed successfully across the one-, two-, and four-worker configurations. Wireshark captures confirmed TLS 1.3 handshake messages and encrypted application data exchanged between the master and ESP32-S3 workers during distributed image processing, as illustrated in Figure 4.

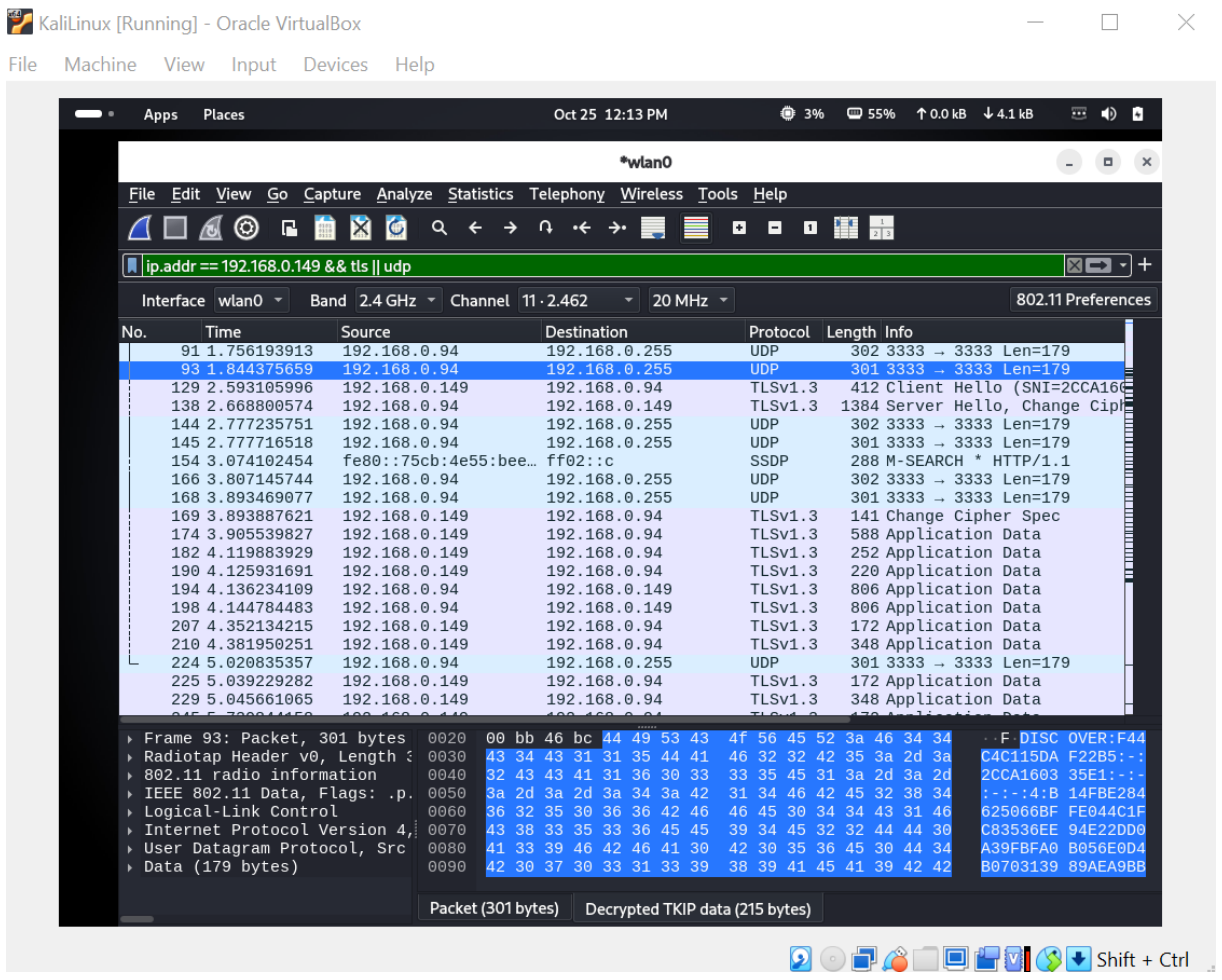

*Figure 4. Wireshark capture of UDP discovery and TLS 1.3 encrypted communication.*

A controlled private-key replacement experiment was conducted to evaluate TLS authentication enforcement. After replacing an ESP32-S3 worker's private key, the master repeatedly rejected the TLS handshake with a BAD_SIGNATURE error. The worker subsequently failed to transmit its ALIVE message, preventing successful worker admission, as shown in Figure 5.

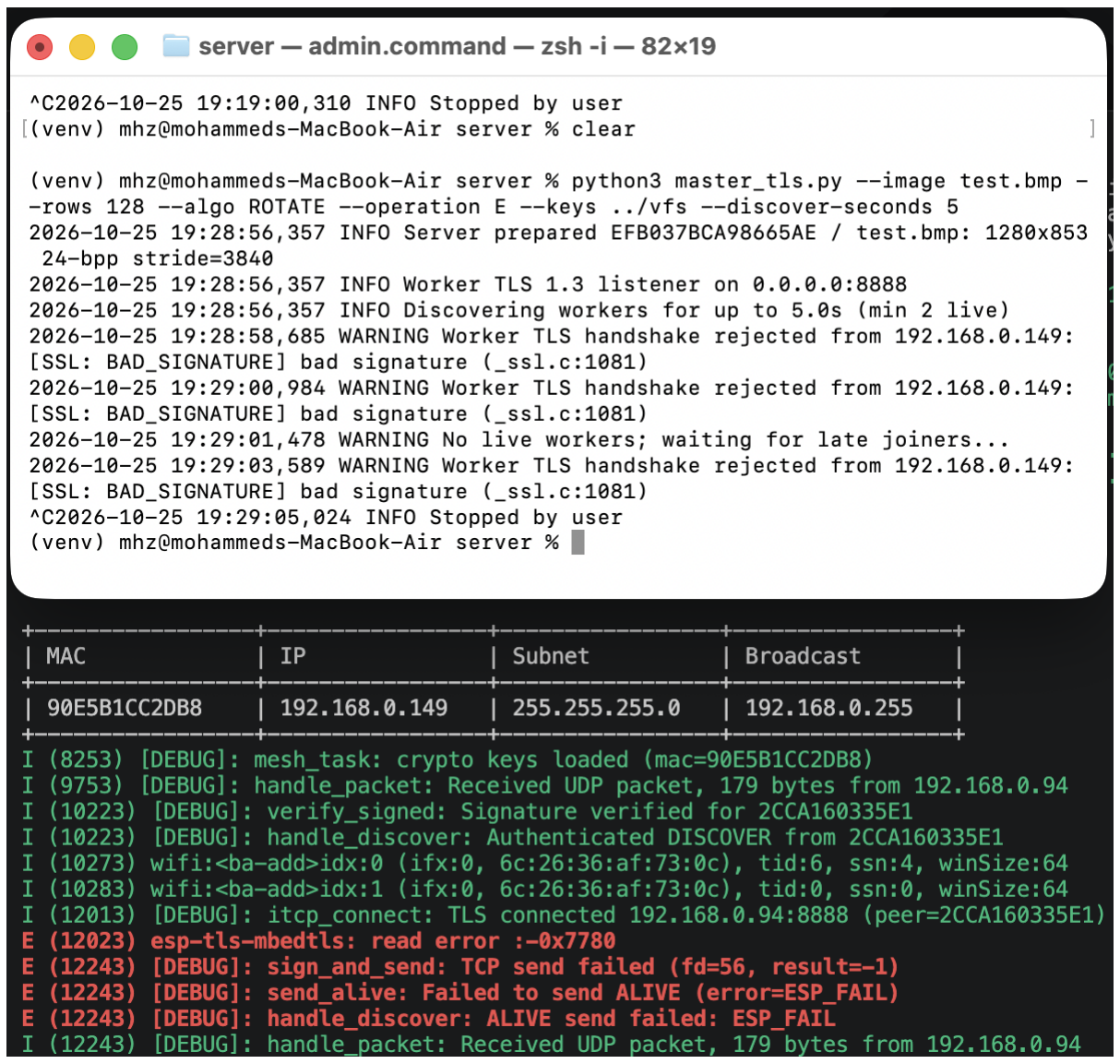

*Figure 5. TLS 1.3 authentication failure after certificate replacement.*

These observations confirm successful TLS 1.3 encrypted communication and effective cryptographic authentication, demonstrated by the rejection of a mismatched private-key configuration. Together, these results support the reliability of the implemented mutual authentication and secure communication mechanisms.

## 5.4 Worker Flexibility and Image Evaluation

The experiments evaluated the flexibility of the implemented master-coordinated chaotic image-processing architecture by encrypting and decrypting the same image using different ESP32-S3 worker configurations and block sizes. The experiment used the SUBSTITUTE algorithm with TLS 1.3 communication. Table 4 summarizes the configurations and execution results.

TABLE 4
WORKER CONFIGURATION AND EXECUTION RESULTS

| Parameter | Encryption | Decryption |
|---|---|---|
| ESP32-S3 workers | 4 | 2 |
| Rows per task | 128 | 256 |
| Total tasks | 7 | 4 |
| Execution time (s) | 15.109 | 18.064 |

The results demonstrate successful encryption and decryption using different worker counts and block sizes. This confirms that the master preserves the required chaotic sequence independently of worker assignments and task partitions, enabling correct image recovery. Figures 6 and 7 present the corresponding master execution logs, confirming successful completion of all assigned tasks.

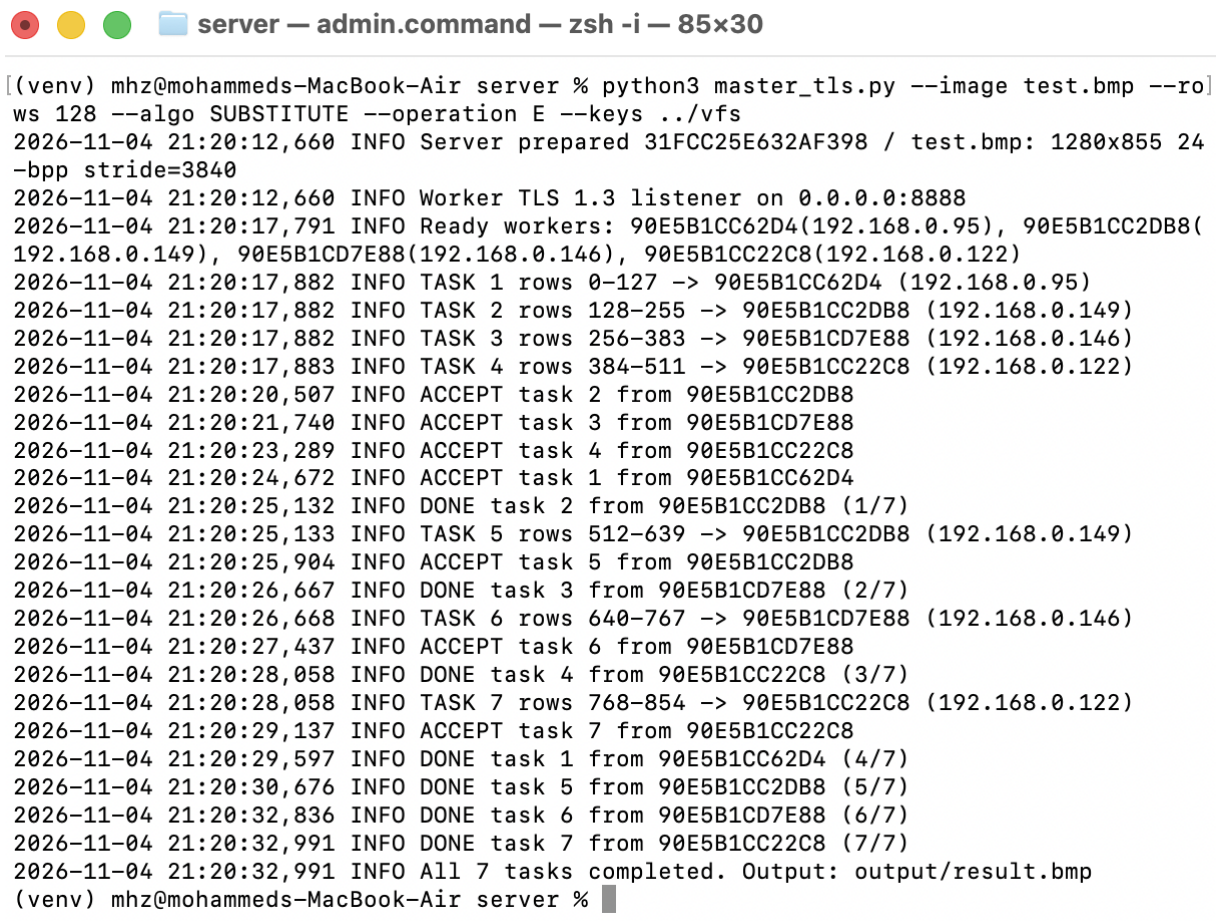

*Figure 6. Encryption using four ESP32-S3 workers with 128-row blocks.*

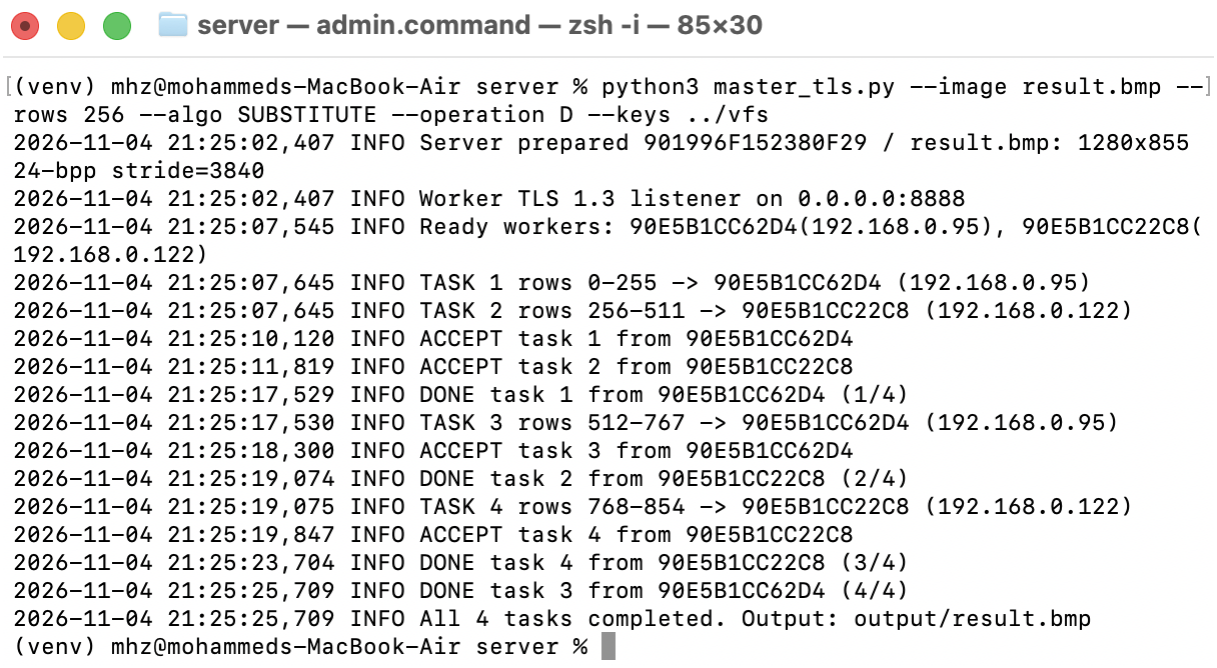

*Figure 7. Decryption using two ESP32-S3 workers with 256-row blocks.*

Image correctness and statistical characteristics were evaluated using SHA-256, Shannon entropy, Structural Similarity Index (SSIM), and adjacent-pixel correlation. The results are summarized in Table 5.

TABLE 5
IMAGE CORRECTNESS AND STATISTICAL RESULTS

| Evaluation | Result | Interpretation |
|---|---|---|
| Shannon entropy | 7.860255 → 7.999940 bits/byte | Near maximum (8 bits/byte) |
| SHA-256 | 4640051beba73f44a53fc4b4 d0544cd0ac0c422f45eb7c45 78ee588c1ca7187c | Original = Decrypted (exact recovery) |
| SSIM (Original vs. Encrypted) | 0.008762 | Very low similarity |
| SSIM (Original vs. Decrypted) | 1.000000 | Perfect similarity |
| Horizontal correlation | −0.0173 to 0.0227 | Near zero |
| Vertical correlation | −0.0261 to 0.0106 | Near zero |
| Diagonal correlation | −0.0245 to 0.0257 | Near zero |

Shannon entropy increased after encryption, approaching the theoretical maximum and indicating a more uniform distribution of pixel values. The near-zero adjacent-pixel correlations and low SSIM between the original and encrypted images further indicate reduced statistical similarity. Matching SHA-256 hashes and an SSIM of 1.000000 confirm exact image recovery after decryption.

However, these statistical results alone do not establish encryption security. The evaluation results are summarized in Table 5.

Figures 8–10 illustrate image recovery, RGB intensity distributions, and adjacent-pixel correlation, respectively. The corresponding experimental verification logs are provided in Appendix B.

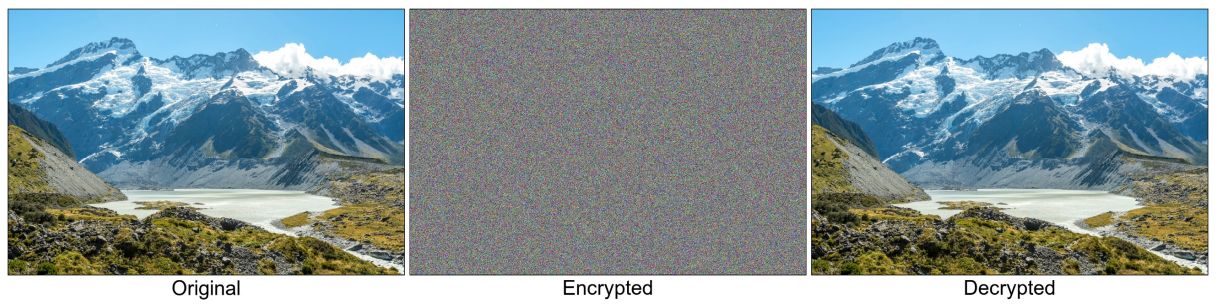

*Figure 8. Original, encrypted, and decrypted images.*

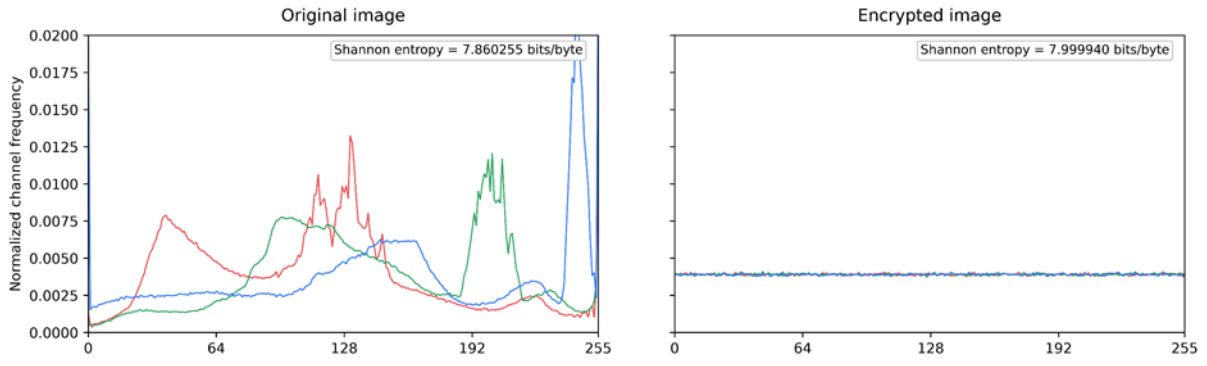

*Figure 9. RGB intensity distributions of the original and encrypted images.*

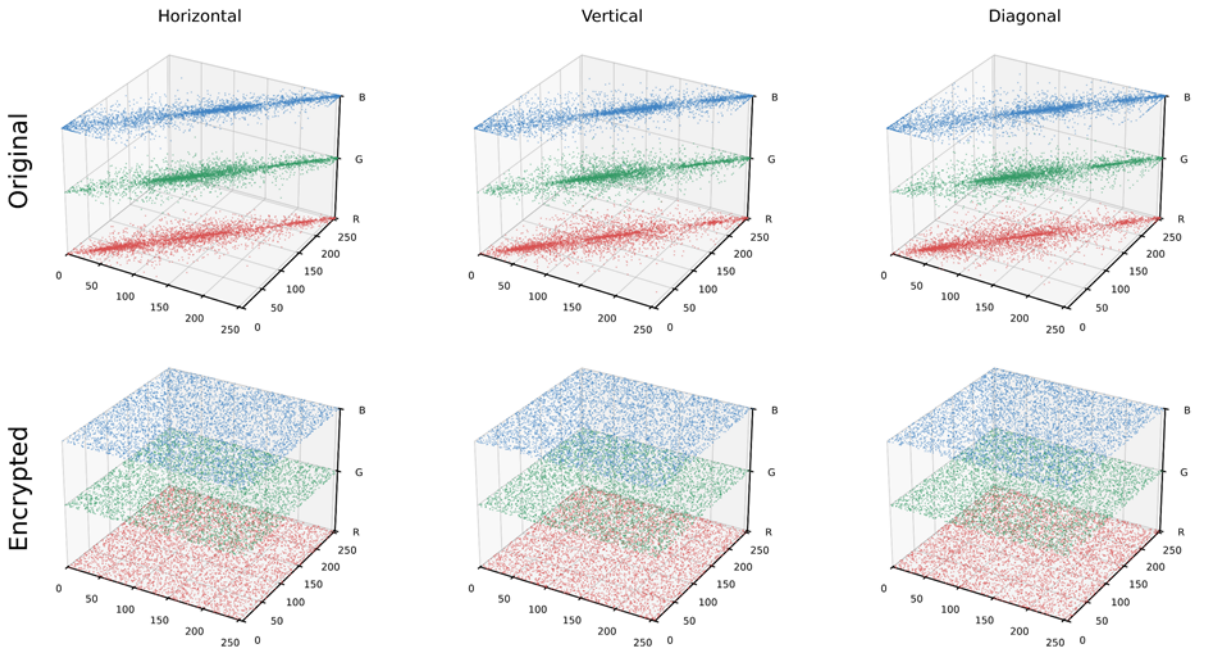

*Figure 10. Adjacent-pixel correlation analysis.*

The results confirm lossless image recovery and substantial statistical differences between original and encrypted images. The experiment also demonstrates that the master maintains chaotic-sequence continuity across different worker assignments and task partitions, preserving image correctness.

### 5.5 Discussion

The sublinear speed-up suggests that communication and coordination overhead may limit scalability. However, assigning chaotic-sequence segments according to image-block positions rather than worker identities allows task reassignment without disrupting sequence continuity.

## 6. Conclusion

This study designed and evaluated an authenticated distributed chaotic image-processing system using independent Wi-Fi-connected ESP32-S3 microcontrollers. The experimental results demonstrated improved processing performance, lossless image recovery, flexible worker configurations, and successful task reassignment following worker interruptions. ECDSA authentication and TLS 1.3 provided secure communication between the master and workers. Future work may explore additional chaotic algorithms, larger image datasets, and further scalability improvements.

## 7. References

[1] H. B. Prajapati and S. K. Vij, “Analytical study of parallel and distributed image processing,” in 2011 International Conference on Image Information Processing, Nov. 2011, pp. 1–6. doi: 10.1109/ICIIP.2011.6108870.

[2] S. Kumaran and M. J. Quinn, “Divide-and-conquer programming on MIMD computers,” in Proceedings of 9th International Parallel Processing Symposium, Apr. 1995, pp. 734–741. doi: 10.1109/IPPS.1995.395867.

[3] C. Nicolescu and P. Jonker, “EASY-PIPE - An ‘Easy to use’ parallel image processing environment based on algorithmic skeletons,” in Proceedings 15th International Parallel and Distributed Processing

Symposium, Apr. 2001, pp. 1151–1157. doi: 10.1109/IPDPS.2001.925086.

[4] P. Chalermwat, N. Alexandridis, P. Piamsa-Nga, and M. O'Connell, "Parallel image processing in heterogeneous computing network systems," in Proceedings of 3rd IEEE International Conference on Image Processing, Sep. 1996, pp. 161–164, vol. 2. doi: 10.1109/ICIP.1996.560627.

[5] M. Dongdong, L. Jinzong, Z. Bing, and Z. Fuzhen, "Research on the Architectures of Parallel Image Processing Systems," in 2008 Second International Symposium on Intelligent Information Technology Application, Dec. 2008, pp. 146–150. doi: 10.1109/IITA.2008.182.

[6] F. Niederl and A. Goller, "Method execution on a distributed image processing back-end," in Proceedings of the Sixth Euromicro Workshop on Parallel and Distributed Processing, Jan. 1998, pp. 243–249. doi: 10.1109/EMPDP.1998.647205.

[7] T. Kruger, J. Wickel, and K. F. Kraiss, "Parallel and distributed computing for an adaptive visual object retrieval system," in Proceedings International Parallel and Distributed Processing Symposium, Apr. 2003. doi: 10.1109/IPDPS.2003.1213423.

[8] N. Khayati, W. Lejouad-Chaari, and S. Sevestre-Ghalila, "A distributed image processing support system application to medical imaging," in 2008 IEEE International Workshop on Imaging Systems and Techniques, Sep. 2008, pp. 261–264. doi: 10.1109/IST.2008.4659981.

[9] B. Rinner, M. Jovanovic, and M. Quaritsch, "Embedded Middleware on Distributed Smart Cameras," in 2007 IEEE International Conference on Acoustics, Speech and Signal Processing, Apr. 2007, pp. IV-1381–IV-1384. doi: 10.1109/ICASSP.2007.367336.

[10] L. Dong et al., "A Hierarchical Distributed Processing Framework for Big Image Data," IEEE Transactions on Big Data, vol. 2, no. 4, pp. 297–309, Dec. 2016. doi: 10.1109/TBDATA.2016.2613992.

[11] M. Rakhimov, D. Mamadjanov, and A. Mukhiddinov, "A High-Performance Parallel Approach to Image Processing in Distributed Computing," in 2020 IEEE 14th International Conference on Application of Information and Communication Technologies, Oct. 2020, pp. 1–5. doi: 10.1109/AICT50176.2020.9368840.

[12] C. Li, X. Cao, W. Ren, J. Yuan, and X. Cao, "An Ultra High Definition Array Camera System and Parallel Processing Technology," in 2024 Asia-Pacific Conference on Image Processing, Electronics and Computers, Apr. 2024, pp. 239–244. doi: 10.1109/IPEC61310.2024.00048.

[13] R. S. M and R. S. Raj, "Analysis of Edge Detection Technique over Parallel Environment," in 2024 Second International Conference on Advances in Information Technology, Jul. 2024, pp. 1–6. doi: 10.1109/ICAIT61638.2024.10690341.

[14] R. C. Ivanescu, R. Nagy, A. I. Ofiteru, C. Comanescu, and A. Manda, "Parallel vs Distributed Edge Detection for Large Medical Image Datasets," in 2022 E-Health and Bioengineering Conference, Nov. 2022, pp. 1–4. doi: 10.1109/EHB55594.2022.9991599.

[15] J. Kim and D. Nam, "Enhanced JPEG Decoding Using PIM Architectures with Parallel MCU Processing," in 2025 IEEE International Parallel and Distributed Processing Symposium, Jun. 2025, pp. 1228–1237. doi: 10.1109/IPDPS64566.2025.00111.

[16] P. Bagul and V. Inamdar, "Hardware Up-gradation Plugin for Accelerating the Image Convolution Processing," in 2022 IEEE Region 10 Symposium, Jul. 2022, pp. 1–5. doi: 10.1109/TENSYMP54529.2022.9864472.

[17] R. Rajak, D. Raveendran, M. C. Bh, and S. S. Medasani, "High Resolution Satellite Image Processing Using Hadoop Framework," in 2015 IEEE International Conference on Cloud Computing in Emerging Markets, Nov. 2015, pp. 16–21. doi: 10.1109/CCEM.2015.16.

[18] R. A. Brown, H. Zhu, and J. R. Mitchell, "Distributed vector Processing of a new local MultiScale Fourier transform for medical imaging applications," IEEE Transactions on Medical Imaging, vol. 24, no. 5, pp. 689–691, May 2005. doi: 10.1109/TMI.2005.845320.

[19] J. Tang, A. Zhang, and S. Li, "Preliminary Research on Grid-based Remote Sensing Image Distributed Processing," in 2007 IFIP International Conference on Network and Parallel Computing Workshops, Sep. 2007, pp. 408–412. doi: 10.1109/NPC.2007.54.

[20] Y. Hu, Q. Xi, L. Xiao, Q. Fang, and Y. Hu, "Research on Image Analysis and Processing Technology Based on Big Data Technology," in 2023 IEEE 2nd International Conference on Electrical Engineering, Big Data and Algorithms, Feb. 2023, pp. 1383–1386. doi: 10.1109/EEBDA56825.2023.10090749.

[21] V. Hristov, M. Doichev, A. Ismailov, and D. Pepedzhiev, "A Web-Based Image Segmentation Tool for Educational Purposes Using Java," in 2025 9th International Symposium on Multidisciplinary Studies and Innovative Technologies, Nov. 2025, pp. 1–6. doi: 10.1109/ISMSIT67332.2025.11267869.

[22] U. S. N. Raju, K. S. Kumar, V. Mehta, R. Sharma, and S. Kuli, "Cluster based block processing for gigantic images: Dimension and size," in 2017 Fourth International Conference on Image Information Processing, Dec. 2017, pp. 1–5. doi: 10.1109/ICIIP.2017.8313778.

[23] J. Che, Y.-S. Shi, Y. Xiang, and Y.-T. Ma, “The fractional differential enhancement of image texture features and its parallel processing optimization,” in 2012 5th International Congress on Image and Signal Processing, Oct. 2012, pp. 330–333. doi: 10.1109/CISP.2012.6470034.

[24] B. Ge, Z. Shen, and J. Zhang, “Fast Chaotic Image Encryption Algorithm Using a Novel Divide and Conquer Diffusion Strategy,” IEEE Access, vol. 10, pp. 95986–96005, 2022. doi: 10.1109/ACCESS.2022.3204873.

[25] J. Wang, Y.-Q. Wang, and Z. Zhang, “A self-adaptive image cryptosystem based on hyper-chaos,” in 2016 35th Chinese Control Conference, Jul. 2016, pp. 2385–2390. doi: 10.1109/ChiCC.2016.7553721.

[26] P. Ping, X. Zhang, X. Yang, Y. Mao, and Z. Gao, “Parallel Image Encryption Technology Based on Cellular Automaton,” in 2020 IEEE Sixth International Conference on Big Data Computing Service and Applications, Aug. 2020, pp. 216–223. doi: 10.1109/BigDataService49289.2020.00042.

[27] G. Gambhir, J. K. Mandal, and M. Gambhir, “Parallel Chaos Hash based LSB Steganography Technique using Logistic Map,” in 2022 Seventh International Conference on Parallel, Distributed and Grid Computing, Nov. 2022, pp. 308–312. doi: 10.1109/PDGC56933.2022.10053352.

[28] F. Zhang, “Protection of Express Information Using Parallel Computing,” in 2023 4th International Conference on Information Science, Parallel and Distributed Systems, Jul. 2023, pp. 188–191. doi: 10.1109/ISPDS58840.2023.10235369.

[29] M. S. Khan, A. Al-Dubai, J. Ahmad, N. Pitropakis, and B. Ghaleb, “A Novel Feature-Aware Chaotic Image Encryption Scheme For Data Security and Privacy in IoT and Edge Networks,” in 2025 International Joint Conference on Neural Networks, Jun. 2025, pp. 1–7. doi: 10.1109/IJCNN64981.2025.11227573.

[30] L. Li, G. Wen, Z. Wang, and Y. Yang, “Efficient and Secure Image Communication System Based on Compressed Sensing for IoT Monitoring Applications,” IEEE Transactions on Multimedia, vol. 22, no. 1, pp. 82–95, Jan. 2020. doi: 10.1109/TMM.2019.2923111.

[31] P. Saraswat, S. Chaudhari, and J. Kundale, “Fractional Discrete Cosine Transform and Chaotic Based Encryption for Medical Image Security,” in 2024 IEEE International Conference on Blockchain and Distributed Systems Security, Oct. 2024, pp. 1–6. doi: 10.1109/ICBDS61829.2024.10837104.

[32] G. Du, “Security of Encrypted Images in Network Transmission Based on an Improved Chaos Algorithm,” Journal of Cyber Security and Mobility, vol. 12, no. 5, pp. 675–696, Sep. 2023. doi: 10.13052/jcsm2245-1439.1253.

[33] M. Abutaha, I. Amar, and S. AlQahtani, “Parallel and Practical Approach of Efficient Image Chaotic Encryption Based on Message Passing Interface (MPI),” Entropy, vol. 24, no. 4, p. 566, Apr. 2022. doi: 10.3390/e24040566.

[34] A. Flores-Vergara et al., “Implementing a Chaotic Cryptosystem by Performing Parallel Computing on Embedded Systems with Multiprocessors,” Entropy, vol. 21, no. 3, p. 268, Mar. 2019. doi: 10.3390/e21030268.

[35] P. Cao, L. Teng, H. H. C. Iu, and Y. Liu, “A Distributed Cloud Storage and Medical Image Encryption Algorithm Application: Medical Image Sharing System,” IEEE Transactions on Multimedia, vol. 28, pp. 3409–3421, 2026. doi: 10.1109/TMM.2026.3651104.

[36] M. I. Wade, M. Chouikha, T. Gill, W. Patterson, T. M. Washington, and J. Zeng, “Distributed Image Encryption Based On a Homomorphic Cryptographic Approach,” in 2019 IEEE 10th Annual Ubiquitous Computing, Electronics & Mobile Communication Conference, Oct. 2019, pp. 686–696. doi: 10.1109/UEMCON47517.2019.8993025.

[37] R. T. Dinesh Kumar, M. S, and G. K, “Secure Medical Record Storage in Cloud using Deep Encryption,” in 2026 International Conference on Electronics and Renewable Systems, Feb. 2026, pp. 1437–1441. doi: 10.1109/ICEARS67481.2026.11416700.

[38] A. Bencherqui, H. Tahiri, M. Tamimi, M. Kmich, M. A. Tahiri, and M. Sayyouri, “Design and Implementation of a Parallel Secure Medical Image Compression-Encryption Scheme on Embedded Cluster Systems,” in 2026 18th International Conference on Electronics, Computers and Artificial Intelligence, Jul. 2026, pp. 1–6. doi: 10.1109/ECAI69016.2026.11613621.

## 8. Appendices

### A. Original Execution Log

The following log documents the controlled fault-injection experiment, including task execution, timeout detection, and automatic reassignment. All seven tasks were successfully completed despite the deliberate worker interruption.

server — admin.command — zsh -i — 99×33

Task Assigned / Started | Task Timeout (Failed) | Task Completed

```
(venv) mhz@mohammeds-MacBook-Air server % python3 master_tls.py --image original.bmp --rows 128 --algo
SUBSTITUTE --operation E --keys ../vfs
2026-11-05 01:32:45,676 INFO Server prepared E0FF6F68C82B2E99 / original.bmp: 1280x855 24-bpp
2026-11-05 01:32:45,677 INFO Worker TLS 1.3 listener on 0.0.0.0:8888
2026-11-05 01:32:50,757 INFO Ready workers: 90E5B1CC62D4(192.168.0.95), 90E5B1CC2DB8(192.168.0.149),
90E5B1CD7E88(192.168.0.146), 90E5B1CC22C8(192.168.0.122)
2026-11-05 01:32:50,850 INFO TASK 1 rows 0-127 -> 90E5B1CC62D4 (192.168.0.95)
2026-11-05 01:32:50,850 INFO TASK 2 rows 128-255 -> 90E5B1CC2DB8 (192.168.0.149)
2026-11-05 01:32:50,850 INFO TASK 3 rows 256-383 -> 90E5B1CD7E88 (192.168.0.146)
2026-11-05 01:32:50,850 INFO TASK 4 rows 384-511 -> 90E5B1CC22C8 (192.168.0.122)
2026-11-05 01:32:53,475 INFO ACCEPT task 1 from 90E5B1CC62D4
2026-11-05 01:32:55,174 INFO ACCEPT task 3 from 90E5B1CD7E88
2026-11-05 01:32:56,719 INFO ACCEPT task 2 from 90E5B1CC2DB8
2026-11-05 01:32:57,954 INFO DONE task 1 from 90E5B1CC62D4 (1/7)
2026-11-05 01:32:57,954 INFO TASK 5 rows 512-639 -> 90E5B1CC62D4 (192.168.0.95)
2026-11-05 01:32:58,265 INFO ACCEPT task 4 from 90E5B1CC22C8
2026-11-05 01:32:58,726 INFO ACCEPT task 5 from 90E5B1CC62D4
2026-11-05 01:32:59,649 INFO DONE task 3 from 90E5B1CD7E88 (2/7)
2026-11-05 01:32:59,650 INFO TASK 6 rows 640-767 -> 90E5B1CD7E88 (192.168.0.146)
2026-11-05 01:33:00,421 INFO ACCEPT task 6 from 90E5B1CD7E88
2026-11-05 01:33:03,199 INFO DONE task 4 from 90E5B1CC22C8 (3/7)
2026-11-05 01:33:03,200 INFO TASK 7 rows 768-854 -> 90E5B1CC22C8 (192.168.0.122)
2026-11-05 01:33:03,659 INFO DONE task 5 from 90E5B1CC62D4 (4/7)
2026-11-05 01:33:04,120 INFO ACCEPT task 7 from 90E5B1CC22C8
2026-11-05 01:33:05,039 INFO DONE task 6 from 90E5B1CD7E88 (6/7)
2026-11-05 01:33:07,665 INFO DONE task 7 from 90E5B1CC22C8 (6/7)
2026-11-05 01:33:10,920 WARNING Task 2 timed out on 90E5B1CC2DB8; requeue
2026-11-05 01:33:10,921 INFO TASK 2 rows 128-255 -> 90E5B1CC62D4 (192.168.0.95)
2026-11-05 01:33:11,847 INFO ACCEPT task 2 from 90E5B1CC62D4
2026-11-05 01:33:16,650 INFO DONE task 2 from 90E5B1CC62D4 (7/7)
2026-11-05 01:33:16,650 INFO All 7 tasks completed. Output: output/result.bmp
(venv) mhz@mohammeds-MacBook-Air server %
```

### B. Image Statistical Analysis and Correctness Verification

C:\Windows\System32\cmd.exe

```
C:\Users\x\Desktop\test\evaluate>python3 histogram.py --original original.bmp --encrypted encrypted.bmp
Original entropy:  7.860255 bits/byte
Encrypted entropy: 7.999940 bits/byte
Saved: histogram_comparison.png

C:\Users\x\Desktop\test\evaluate>verify_sha256.py --original original.bmp --decrypted decrypted.bmp
Original SHA-256:  4640051beba73f44a53fc4b4d0544cd0ac0c422f45eb7c4578ee588c1ca7187c
Decrypted SHA-256: 4640051beba73f44a53fc4b4d0544cd0ac0c422f45eb7c4578ee588c1ca7187c
Result: IDENTICAL - Lossless recovery

C:\Users\x\Desktop\test\evaluate>python3 correlation.py --original original.bmp --encrypted encrypted.bmp
Original - Horizontal: R=0.9502, G=0.9491, B=0.9716
Original - Vertical: R=0.9149, G=0.9152, B=0.9536
Original - Diagonal: R=0.9018, G=0.9029, B=0.9461
Encrypted - Horizontal: R=-0.0173, G=0.0227, B=0.0114
Encrypted - Vertical: R=-0.0261, G=-0.0167, B=0.0106
Encrypted - Diagonal: R=0.0164, G=0.0257, B=-0.0245
Saved: correlation.png

C:\Users\x\Desktop\test\evaluate>
```

## C. TLS 1.3 Communication Evidence

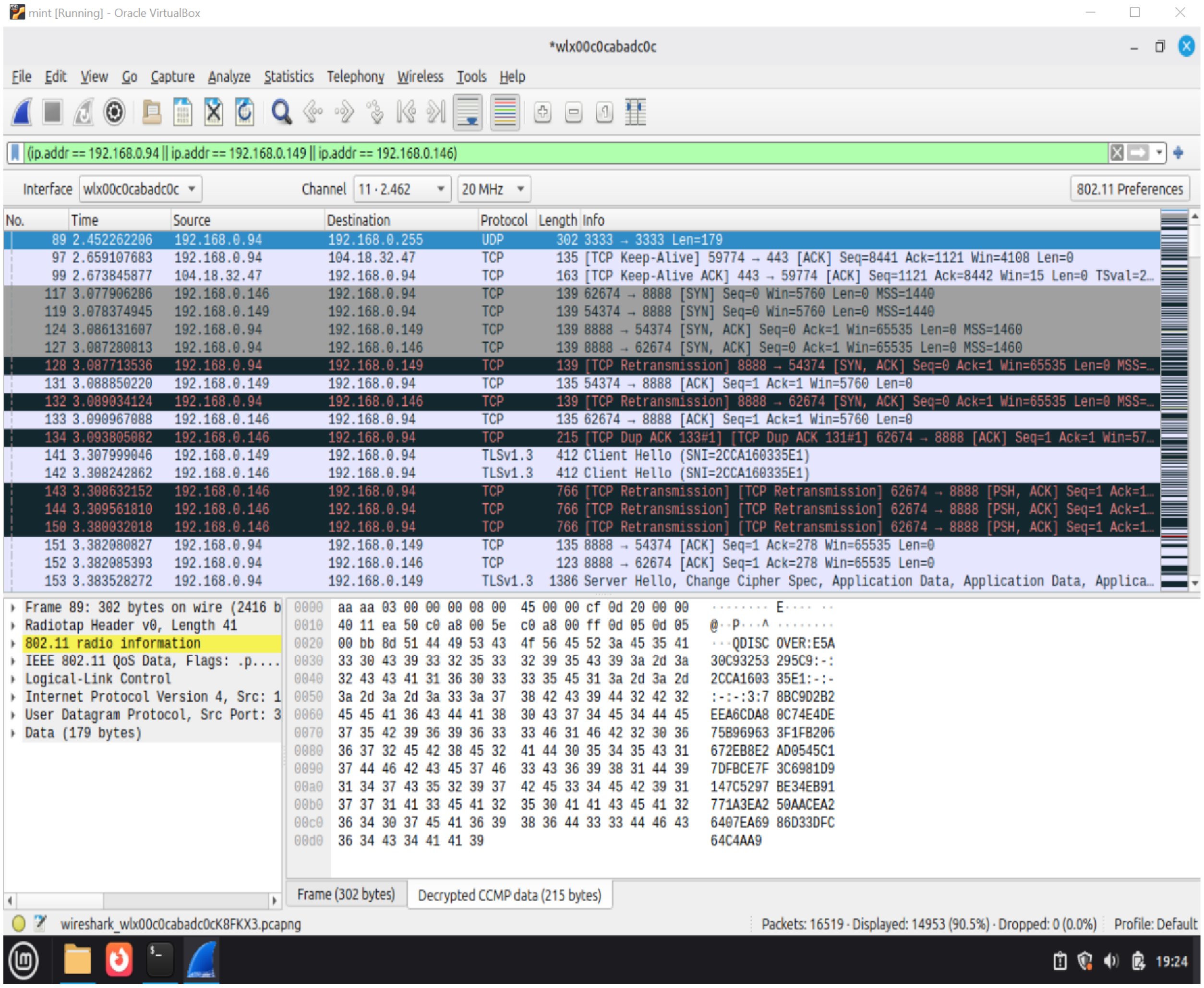

*Figure 11. Wireshark capture of UDP discovery and TLS 1.3 communication.*

**D. Table D1. Feature-Based Comparison of Related Studies**

| Study | Worker / Platform | Multi-node | Chaotic encryption | Coordinated chaotic | Wi-Fi MCUs | ECDSA | TLS 1.3 | Task recovery |
|---|---|---|---|---|---|---|---|---|
| [22] Raju, 2017 | Hadoop computer cluster | ✓ | ✗ | ✗ | ✗ | ✗ | ✗ | ✗ |
| [23] Che, 2012 | MPI computing platform | ✗ | ✗ | ✗ | ✗ | ✗ | ✗ | ✗ |
| [24] Ge, 2022 | Algorithm; hardware unspecified | ✗ | ✓ | ✗ | ✗ | ✗ | ✗ | ✗ |
| [26] Ping, 2020 | CPU/GPU computer | ✗ | ✓ | ✗ | ✗ | ✗ | ✗ | ✗ |
| [28] Zhang, 2023 | Parallel computing platform | ✗ | ✗ | ✗ | ✗ | ✗ | ✗ | ✗ |
| [29] Khan, 2025 | Platform unspecified | ✗ | ✓ | ✗ | ✗ | ✗ | ✗ | ✗ |
| [33] Abutaha, 2022 | Intel i3/i5 multicore computers | ✗ | ✓ | ✗ | ✗ | ✗ | ✗ | ✗ |
| [34] Flores-Vergara, 2019 | 6 × Raspberry Pi 3 | ✓ | ✓ | ✗ | ✗ | ✗ | ✗ | ✗ |
| [37] Kumar, 2026 | Cloud computing nodes | ✓ | ✗ | ✗ | ✗ | ✗ | ✗ | ✗ |
| [38] Bencherqui, 2026 | 4 × Raspberry Pi | ✓ | ✓ | ✗ | ✗ | ✗ | ✗ | ✗ |
| **This work** | 1–4 × ESP32-S3 | ✓ | ✓ | ✓ | ✓ | ✓ | ✓ | ✓ |

***Note***: *✓ = Supported by the study; ✗ = Not reported or demonstrated*